%% file: paper_topq_2012_09.tex
\documentclass[final,3p,times,twocolumn]{elsarticle}

\usepackage{preprintcover}  
\PreprintCoverPaperTitle{Search for single $b^*$-quark production with the ATLAS detector at $\sqrt{s}=7$~TeV}
\PreprintIdNumber{CERN-PH-EP-2012-344}  
\PreprintCoverAbstract{
The results of a search for an excited bottom-quark
$b^*$ in $pp$ collisions at $\sqrt{s}=7\TeV$, using $4.7~\ifb$ of data collected
by the ATLAS detector at the LHC are presented.
In the model studied, a single $b^*$-quark is produced through a
chromomagnetic interaction and subsequently decays to a $W$~boson and a top~quark.
The search is performed in the dilepton and lepton+jets final states, which are combined to set limits 
on $b^*$-quark couplings for a range of $b^*$-quark masses.
For a benchmark with unit size chromomagnetic and Standard Model-like 
electroweak $b^{*}$~couplings,
$b^*$~quarks with masses less than $870\GeV$ are excluded
at the 95\% credibility level.
}  
\PreprintJournalName{Phys. Lett. B}  

\usepackage{amssymb}
\usepackage{lineno}

\graphicspath{{f/}}
\usepackage{atlasphysics}
\usepackage{subfigure}
\usepackage{mathrsfs}
\usepackage{extra_definitions}
\usepackage[hyperindex,breaklinks]{hyperref}

\journal{Phys. Lett. B}

\biboptions{square,comma,numbers,sort&compress}
\begin{document}

\begin{frontmatter}

  \title{Search for single $b^*$-quark production with the ATLAS detector at $\sqrt{s}=7\TeV$}

\author{The ATLAS Collaboration}

\address{}

\begin{abstract}

The results of a search for an excited bottom-quark
$b^*$ in $pp$ collisions at $\sqrt{s}=7\TeV$, using $4.7~\ifb$ of data collected
by the ATLAS detector at the LHC are presented.
In the model studied, a single $b^*$-quark is produced through a
chromomagnetic interaction and subsequently decays to a $W$~boson and a top~quark.
The search is performed in the dilepton and lepton+jets final states, which are combined to set limits 
on $b^*$-quark couplings for a range of $b^*$-quark masses.
For a benchmark with unit size chromomagnetic and Standard Model-like 
electroweak $b^{*}$~couplings,
$b^*$~quarks with masses less than $870\GeV$ are excluded
at the 95\% credibility level.

\end{abstract}


\end{frontmatter}

\section{Introduction}
\label{sec:intro}
The single top-quark signature is sensitive to many models of new 
physics~\cite{Tait1:2000sh}. Single top-quark production in the Standard Model (SM) has been measured
at the LHC in the $t$-channel~\cite{Aad:2012ux,CMS:2012ep} 
and in association with a $W$ boson ($Wt$-channel)~\cite{ATLAS-WtEvidence,CMS:2012yv}. 
Searches for resonant production
of a new particle which decays with a single top-quark have been carried out in the
$s$-channel production of a top~quark together with a
$b$~quark~\cite{Aad:2012ej,CMS:2012sc}. This Letter presents
the first search for a resonance decaying to a single top-quark
and a $W$~boson~\cite{RS:2012}.
Here we consider the production of an excited quark $b^*$
which decays to a single top-quark and a $W$~boson. This is the first search for
excited-quarks coupling to the third generation of fermions.

\begin{figure}[!h!tbp]
  \begin{center}
    \includegraphics[width=0.9\columnwidth]{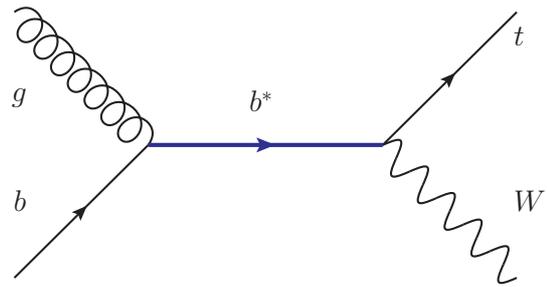}
    \caption{Leading-order Feynman diagram for single-$b^*$-quark production 
     and decay to $Wt$. }
    \label{fig:BpFeynman}
  \end{center}
\end{figure}

Previous searches for excited quarks have focused on their strong 
interactions~\cite{Aad:2011fq,Chatrchyan:2011ns}, as well as their electromagnetic
interactions~\cite{Aaron:2009iz,ATLAS:2011ai}
with SM~quarks. These searches exploit the coupling between the excited quark and
up or down quarks in the proton.
Here the production of excited-quarks coupling primarily to the third generation 
of SM~quarks is investigated. This coupling occurs for example in Randall--Sundrum 
models that address the strong interaction 
sector~\cite{Cheung:2007bu,Fitzpatrick:2007sa}
or in models with a heavy gluon partner, such as composite Higgs 
models~\cite{Bini:2011zb,Vignaroli:2012sf,Vignaroli:2012si}.
The $b^*$~quark is produced singly through its coupling to a 
$b$~quark and a gluon, as shown in Fig.~\ref{fig:BpFeynman}.

The Lagrangian describing this interaction is given by ~\cite{AdeRujula1984253,Baur:1987ga}
\begin{equation}
  {\mathcal L} = \frac{g_s}{2\,\Lambda} G_{\mu\nu}\, 
          \overline{b}\, \sigma^{\mu\nu} 
          \biggl(\kappa^b_L P_L + \kappa^b_R P_R\biggr) b^*
          + \mathrm{h.c.}\, , \label{eq:productionL}
	\end{equation}
where $g_s$ is the strong coupling, $G_{\mu\nu}$ the gauge field tensor of the 
gluon and $\Lambda=m_{b^*}$ the scale of the new physics. 
$P_L$ and $P_R$ are the left- and right-handed
projection operators and $\kappa^b_L$ and $\kappa^b_R$ are the respective coupling
strengths. This analysis is thus complementary to excited-quark searches focusing
on the coupling to the first generation~\cite{:2012ak,Aad:2011yn,Aad:2011fq}.
Single $b^*$-quark production
can also reveal the chiral nature of the excited bottom-quark~\cite{RS:2012}.

In addition to the chromomagnetic coupling,
the $b^*$~quark investigated here also
has weak couplings, as in a general class of new physics
models where new heavy particles stabilise the Higgs-boson mass at the electroweak
scale~\cite{Hill:2002ap,Martin:2009bg,Kumar:2010vx,Holdom:2009rf,Alok:2010zj}.
In such models, the heavy quarks can have left-handed or right-handed
couplings to the $W$~boson or can be vector-like with equal strength for both
couplings. The Lagrangian describing the electroweak decay of the $b^*$~quark,
shown in Fig.~\ref{fig:BpFeynman}, is
\begin{equation}
 {\mathcal L} = \frac{g_2}{\sqrt{2}} \,W^+_{\mu} \,\overline{t} \gamma^{\mu}\, 
                  \bigl(g_L P_L + g_R P_R \bigr) \,b^* + \mathrm{h.c.} \, ,
   \label{eq:decayL}
\end{equation}
where $g_2$ is the $\mathrm{SU(2)}_L$ weak coupling and $g_L$ and $g_R$ are the coupling
strengths for left-handed and right-handed couplings, respectively.

While the search is general and considers any resonance decaying into the $Wt$ signature,
three specific $b^*$-quark coupling scenarios are considered in order to extract 
$b^*$-quark coupling and mass limits: 
left-handed ($\kappa^b_L$, $g_L$ non-zero and $\kappa^b_R=g_R=0$), 
right-handed ($\kappa^b_L=g_L=0$ and $\kappa^b_R$, $g_R$ non-zero) and
vector-like ($\kappa^b_L=\kappa^b_R=\kappa^b_{L/R}$ and $g_L=g_R=g_{L/R}$ non-zero)
production and decay.
Limits are derived as a function of the $b^*$-quark mass as well as the couplings
$\kappa^b_{L,R}$ and $g_{L,R}$. These limits take into account both the change of the
production cross-section and the $b^*\rightarrow Wt$ decay branching ratio, 
which depend on the couplings and the $b^*$-quark mass. The branching ratio to 
$Wt$ varies between 20\% at $m_{b^*}=300~\GeV$ and 40\% at higher values, with decays to $bg$,
$bZ$ and $bH$ also allowed. Contributions from non-$Wt$ decay modes that may
increase the $b^*$-quark acceptance of this analysis are not considered,
resulting in conservative limits.
Signal event yields presented in the following tables are calculated with $\kappa^b_L=g_L=1$ 
and $\kappa^b_R=g_R=0$.

For a left-handed $b^*$ at $\sqrt{s}= 7\TeV$ with $\kappa^b_L=g_L=1$ and $\kappa^b_R=g_R=0$, the leading-order cross section times
branching ratio to $Wt$ is 0.80~pb for $m_{b^*}=900\GeV$~\cite{RS:2012}. 
The uncertainties due to the choice of factorisation and renormalisation scales are evaluated by varying the scales between $m_{b^*}/2$ and $2 \times m_{b^*}$, and those due to the choice of PDF by comparing results obtained using the CT10~\cite{Lai:2010vv}, MRST~\cite{springerlink:10} and NNPDF~\cite{Ball:2010de} sets.  These uncertainties are added in quadrature to yield cross-section uncertainties ranging from 12\% at $m_{b^*}=300\GeV$ to 25\% at $m_{b^*}=1200\GeV$.

This channel proceeds via two $W$~bosons from $b^*$-quark and top-quark decays.
At least one $W$~boson is required to decay to a lepton (electron or muon). 
The analysis is performed separately 
in the dilepton and lepton+jets final states. The lepton+jets channel has the 
advantage that the invariant mass of the $b^*$~quark can be reconstructed, 
whereas the dilepton channel benefits from smaller backgrounds. A discriminant
that separates the $b^*$-quark signal from the backgrounds is defined in each final
state. Limits on $b^*$-quark production are obtained from a combined Bayesian 
analysis of both discriminant distributions.

\section{The ATLAS detector}
The ATLAS detector~\cite{ATL-2008-001} is a general purpose detector with a precise tracking system, calorimeters
and an outer muon spectrometer. The inner tracking system consists of a silicon pixel 
detector, a silicon microstrip tracker, and a straw-tube transition radiation tracker. 
This system is immersed in a $2~\mathrm{T}$ axial magnetic field produced by a solenoid and
provides charged particle tracking and identification in the pseudorapidity\footnote{ATLAS 
uses a right-handed coordinate system with its origin at
the nominal interaction point (IP) in the centre of the detector and the $z$-axis 
along the beam pipe. The $x$-axis points from the IP to the centre
of the LHC ring, and the $y$-axis points upwards. Cylindrical coordinates $(r, \phi)$ 
are used in the transverse plane, $\phi$ is the azimuthal angle around the beam pipe.
The pseudorapidity $\eta$ is defined in terms of the polar angle $\theta$ as
$\eta = -~\rm{ln}(\tan\theta/2)$.} region $|\eta|<2.5$.
The central calorimeter system consists of a liquid-argon electromagnetic sampling
calorimeter with high granularity and an iron/scintillator tile calorimeter providing
hadronic energy measurements in the central pseudorapidity range ($|\eta|<1.7$). 
The endcap and forward regions are instrumented with liquid-argon calorimeters for both
electromagnetic and hadronic energy measurements up to $|\eta|=4.9$.
The muon spectrometer is operated in a toroidal magnetic 
field provided by air-core superconducting magnets and includes tracking chambers 
for precise muon momentum measurements up to $|\eta|=2.7$ and trigger chambers 
covering the range $|\eta|<2.4$.

\section{Data and simulated samples}
\label{sec:DataMC}
 This analysis uses data collected with the ATLAS detector in 2011,
corresponding to an integrated luminosity of
$4.7\pm 0.2~\ifb$~\cite{ATL-LUMI,ATL-CONF-2011-116} of $7\TeV$ proton--proton ($pp$)
collisions delivered by the LHC.
The data are selected using single-electron or single-muon 
triggers whose efficiencies reach their plateau at $25\GeV$ and $20\GeV$,
respectively~\cite{Aad:2012xs,Hristova:1485638}. 
The data must also pass stringent quality requirements~\cite{ATLAS-JETCLEAN}.
Events are selected if they contain at least one primary vertex candidate
with at least five associated tracks.

The signal is modelled using 
{\sc MadGraph5}~\cite{Alwall:2011uj} and the {\sc CTEQ6L1} parton distribution 
functions (PDFs)~\cite{Nadolsky:2008zw}. Events with single top-quarks in the $t$-channel are 
generated with the {\sc AcerMC}~\cite{SAMPLES-ACER} generator,
using the MRST~LO** PDF set~\cite{Sherstnev:2007nd}. {\sc MadGraph5} and
{\sc AcerMC} are interfaced to {\sc Pythia}~\cite{SAMPLES-PYTHIA} for parton showering and modelling
of the underlying event.
Other processes producing single top-quarks and top-quark pairs ($\ttbar$) are modelled
with the next-to-leading-order(NLO) generator {\sc MC@NLO}~\cite{SAMPLES-MCNLO} using the
{\sc CT10} PDF set~\cite{Lai:2010vv}, interfaced to {\sc Herwig}~\cite{SAMPLES-HERWIG} for parton
showering and {\sc Jimmy}~\cite{SAMPLES-JIMMY} for the underlying event.
{\sc Alpgen}~\cite{SAMPLES-ALPGEN} is used to model vector boson ($W$ and $Z$) 
production in association with jets as well as diboson processes ($WW$, $WZ$ and $ZZ$)
using the {\sc CTEQ6L1} PDF set. It is interfaced to {\sc Herwig} for parton shower 
modelling.
In the lepton+jets analysis the diboson processes are modelled with {\sc Herwig} only.
Decays of $\tau$ leptons are handled by {\sc Tauola}~\cite{Davidson:2010rw}.
A top-quark mass of $172.5\GeV$~\cite{PDG} is assumed.
Approximate next-to-next-to-leading-order(NNLO) cross-section calculations are used to normalise the \ttbar~\cite{Aliev:2010zk} ({\sc Hathor})
and single top-quark samples~\cite{Kidonakis:2011wy,Kidonakis:2010tc,Kidonakis:2010ux}, 
while the vector boson and diboson samples are normalised using calculations with {\sc MCFM}~\cite{Campbell:2010ff}
at NNLO and NLO, respectively.

A variable number of additional $pp$ interactions (pile-up) are
overlaid on simulated events, which are then weighted to reproduce the distribution
of the number of collisions per bunch crossing observed in data.
All samples are passed through a {\sc GEANT4}-based simulation~\cite{SAMPLES-G4} of 
the ATLAS detector~\cite{ATL-2010-005_sgtop} and are then reconstructed using the same 
procedure as for collision data.

\section{Physics object selection}
\label{sec:selection}

Electron candidates are reconstructed from
clusters of energy deposits in the calorimeter~\cite{ATL-ELEC-PERF}. 
The transverse energy $E_\mathrm{T}$ of electron candidates is required to be larger than 
$25\GeV$ and their pseudorapidity is required to be $|\eta|<2.47$. 
Electrons in the barrel--endcap transition region of the calorimeter, corresponding
to $1.37<|\eta|<1.52$, are not considered. Selected electrons must pass a set of
``tight'' quality criteria~\cite{ATL-ELEC-PERF} and the electons must be matched to
a track reconstructed in the inner tracking system. Electrons must also be isolated
from close-by tracks in a cone of $\Delta R =\sqrt{(\Delta\eta)^2+(\Delta\phi)^2} <0.3$ 
and from calorimeter energy deposits not belonging to the electron candidate in a cone of $\Delta R < 0.2$. 
The isolation requirements on the sum of transverse momenta of tracks in the cone and on the sum of energy deposits 
in the calorimeter in the cone are chosen as a function of $\pt$ and $\eta$ such that an efficiency of 90\% 
for electrons in the simulation is achieved. 

Muon candidates are reconstructed from matching tracks in the muon spectrometer and inner tracking system.
Muons are required to have transverse momentum $\pt > 25\GeV$ 
and $|\eta|<2.5$ and fulfil tight quality criteria~\cite{ATL-2010-012-2}. Muons 
must be isolated from close-by tracks in a cone of $\Delta R <0.3$
and from energy deposits in the calorimeter in a cone of $\Delta R < 0.2$.
The sum of transverse momenta of tracks in the cone must not exceed $2.5\GeV$ and the sum of energy deposits in the calorimeter in the cone must be below $4\GeV$.

In order to reject events in which a muon emitting a hard photon is also reconstructed 
as an electron, events are vetoed if a selected electron--muon pair shares the same 
track.

Jets are reconstructed from clusters of energy deposits in the
calorimeter~\cite{ATLAS-TOPO-CLUSTER}
using the anti-$k_{t}$ algorithm~\cite{Salam:2008} with a radius parameter $R=0.4$.
These jets are calibrated to the hadronic energy scale
through $\pT$- and $\eta$-dependent scale factors, which are derived from simulation.
An additional uncertainty due to residual differences between simulation and data
is applied in the analysis~\cite{ATLAS-JES-2011}.
Jets are required to have $\pt > 30~(25) \GeV$ and $|\eta| < 2.5$ in the 
dilepton (lepton+jets) channel.
The ratio of the scalar sum of the $\pt$ of tracks associated with the jet and the 
primary vertex to the scalar sum of the $\pt$ of all tracks associated with the jet
must be at least 0.75 to reject jets from pile-up interactions.
Muons overlapping with jets within $\Delta R<0.4$ are removed and the jet is kept.
The closest jet overlapping with electrons within $\Delta R <0.2$
is removed and the electron is kept. If electrons subsequently still overlap with
any remaining jet within $\Delta R< 0.4$, they are removed.
Information about jets containing $b$~quarks~\cite{ATLAS-BTAG} is also used
in the lepton+jets channel. A neural network combines lifetime-related
information reconstructed from the tracks associated with each jet. 
At the chosen working point, the $b$-tagging algorithm has an efficiency of
70\% (20\%/0.7\%) for jets containing $b$~quarks ($c$~quarks/light quarks or gluons)
in a simulated $t\bar{t}$ sample. 

The missing transverse momentum $\met$ is calculated using topological clusters of energy deposits in the
calorimeter and corrected for the presence of muons~\cite{ATLAS-MET}.

\section{Event selection in the dilepton channel}
\label{sec:dilepton}
The event selection and background modelling in the dilepton channel is the same as in the
ATLAS measurement of the single top-quark production in the $Wt$-channel~\cite{ATLAS-WtEvidence}. 
Candidate events must contain exactly two leptons ($ee$, $\mu\mu$ or $e\mu$) with
opposite electric charge and exactly one jet. At least one
of the leptons in each event must match the corresponding trigger-level object.
No $b$-tagging requirement is made since the dominant background from $\ttbar$ 
production also contains $b$~quarks. The $\met$ is required to be 
greater than $50 \GeV$. In the $ee$ and $\mu\mu$ channels, the invariant mass of the
lepton pair, $m_{\ell\ell}$, is required to be outside the $Z$~boson mass window:
$m_{\ell\ell}<81 \GeV$ or $m_{\ell\ell}>101 \GeV$.
In all three channels, the \Ztt\ background is reduced by a dedicated veto, which
requires the sum of the azimuthal angle differences between each lepton and the 
$\met$ vector to be greater than $2.5~\mathrm{rad}$. After all cuts, the acceptance for
signal events with $m_{b^*}=800\GeV$ in which both $W$~bosons decay leptonically 
(to either $e$ or $\mu$) is 26\%.

The main background, accounting for 63\% of the total, comes from $\ttbar$
events in which one of the two jets originating from $b$~quarks is not detected. 
The second largest background is 
from SM $Wt$ production, which has the same final state as the $b^*$-quark signal, and
accounts for 13\% of the total background. Diboson events produced in association with
jets account for 12\% of the total background.
With the exception of single- and diboson samples, these backgrounds are taken from NLO simulation and 
are normalised to their NNLO theoretical predictions. 
Drell--Yan (DY) events contribute a small background of 7.3\% to the sum of 
$ee$ and $\mu\mu$ channel events. The events are taken from the simulation and normalised
to data using a
two-dimensional sideband region with low $\met$ and/or $m_{\ell\ell}$ outside of the $Z$~boson mass window~\cite{ATLAS-WtEvidence}.
The contribution from $\tau\tau$ final states, where both $\tau$ leptons decay
leptonically, is estimated from simulated samples, with the normalisation checked
in an orthogonal data sample obtained by reversing the $Z\to\tautau$ veto cut
described above. $Z\to\tautau$ events account for 0.7\% of the total background.
The small background from jets that are misidentified as primary leptons and from non-prompt leptons (fake dileptons)
is modelled 
and normalised using data~\cite{Aad:2010ey}. 
It accounts for 4\% of the background.

The predicted event yields for the backgrounds and signal at a few mass
points are compared to data in Table~\ref{tab:dilepyields}.
The $\pt$~distributions of the two leptons and the jet are shown in
Fig.~\ref{fig:kinematicsLL1}. 

\begin{table}[htdp]
\begin{center}
 \caption{Observed and predicted event yields in the dilepton channel
with normalisation uncertainties. The signal yields are calculated with 
$\kappa^b_L=g_L=1$ and $\kappa^b_R=g_R=0$.}
\label{tab:dilepyields}
   \begin{tabular}{l*{1}{r@{\ $\pm$\ }l}*{1}}
    Process & \multicolumn{2}{c}{Event yield} \\
    \hline \hline
    $b^*$ (400 \GeV)  & 1250 & 170 \\
    $b^*$ (600 \GeV)  & 211  & 32 \\
    $b^*$ (800 \GeV)  & 41   & 8  \\
    $b^*$ (1000 \GeV) & 8.9  & 1.9  \\
    $b^*$ (1200 \GeV) & 2.1  & 0.5  \\
    \hline
    $Wt$                  & 293  &21   \\
    $t\bar{t}$            & 1380 &140  \\
       Diboson            & 255  &63   \\
    $Z\to\ee$                  & 41   &4    \\
    $Z\to\mumu$                  & 118  &12   \\
    $Z\to\tautau$& 14   &9   \\
    Fake dileptons        & 90   &90  \\
    \hline
    Total expected bkg.   & 2190 &180\\
    Total observed        & \multicolumn{2}{c}{ 2259}\\
   \end{tabular}
\end{center}
\end{table}

\begin{figure}
\subfigure[]{
  \includegraphics[width=0.9\columnwidth]{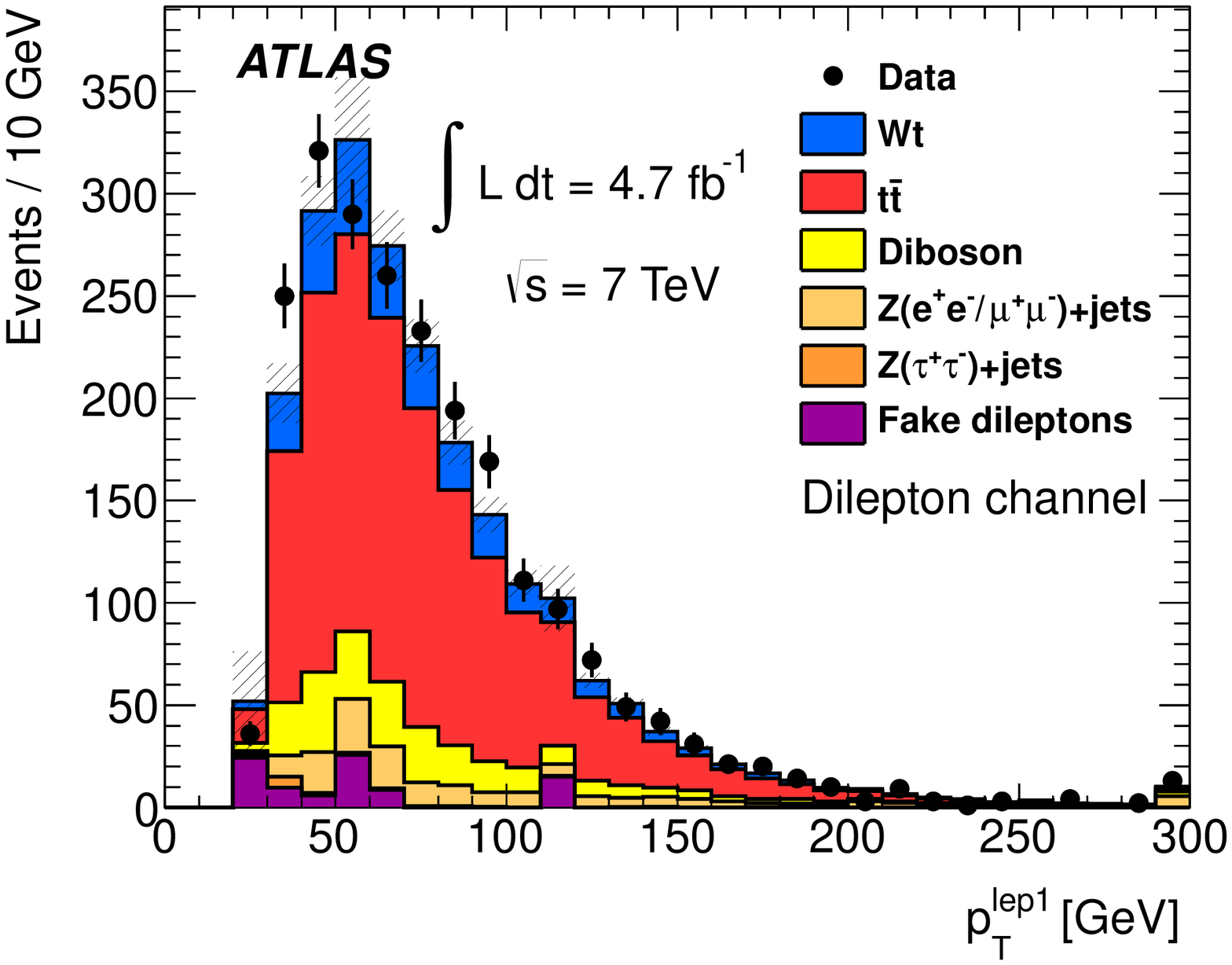}
 \label{fig:ll_lpt}
}
\subfigure[]{
  \includegraphics[width=0.9\columnwidth]{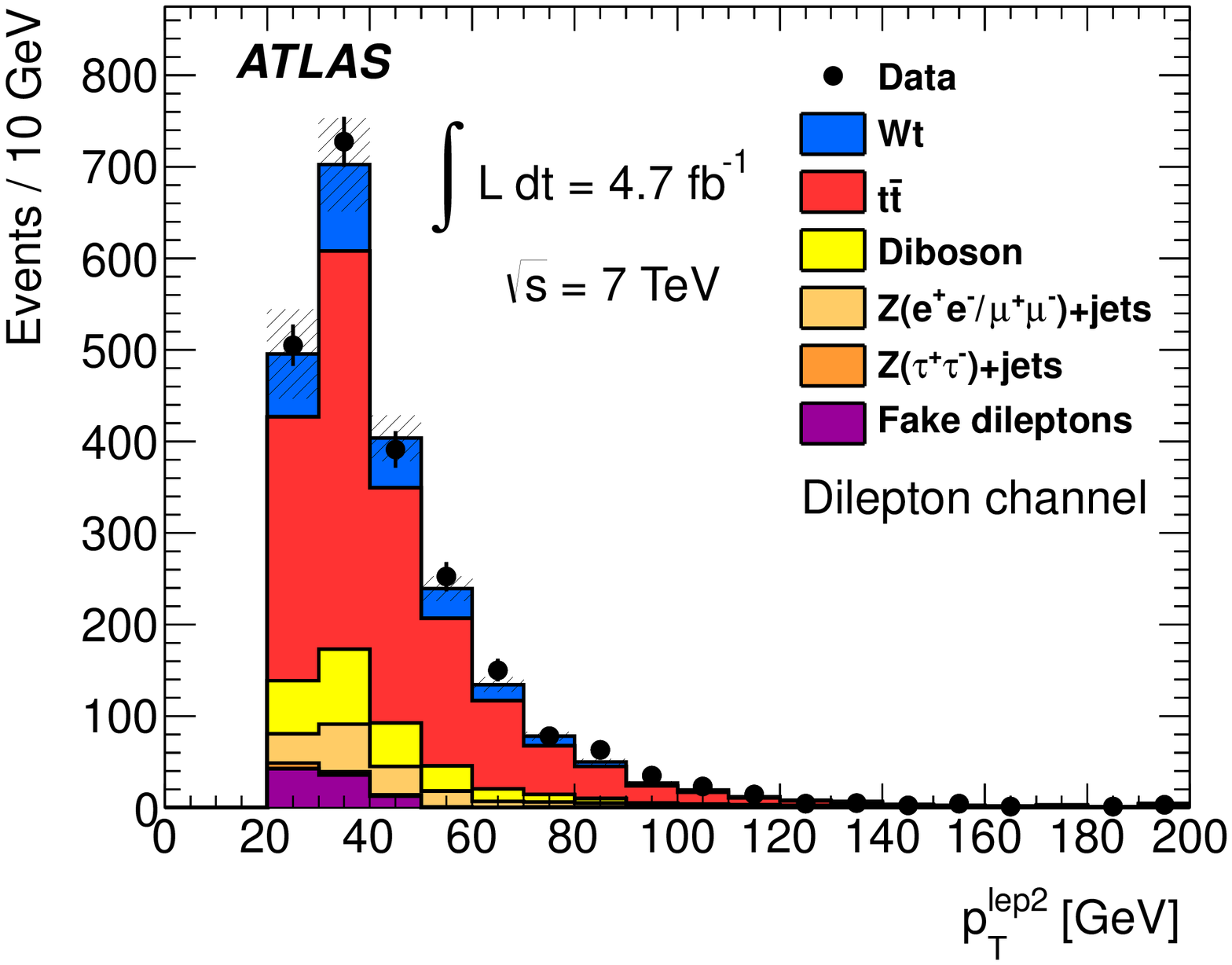}
 \label{fig:ll_lpt2}
}
\subfigure[]{
  \includegraphics[width=0.9\columnwidth]{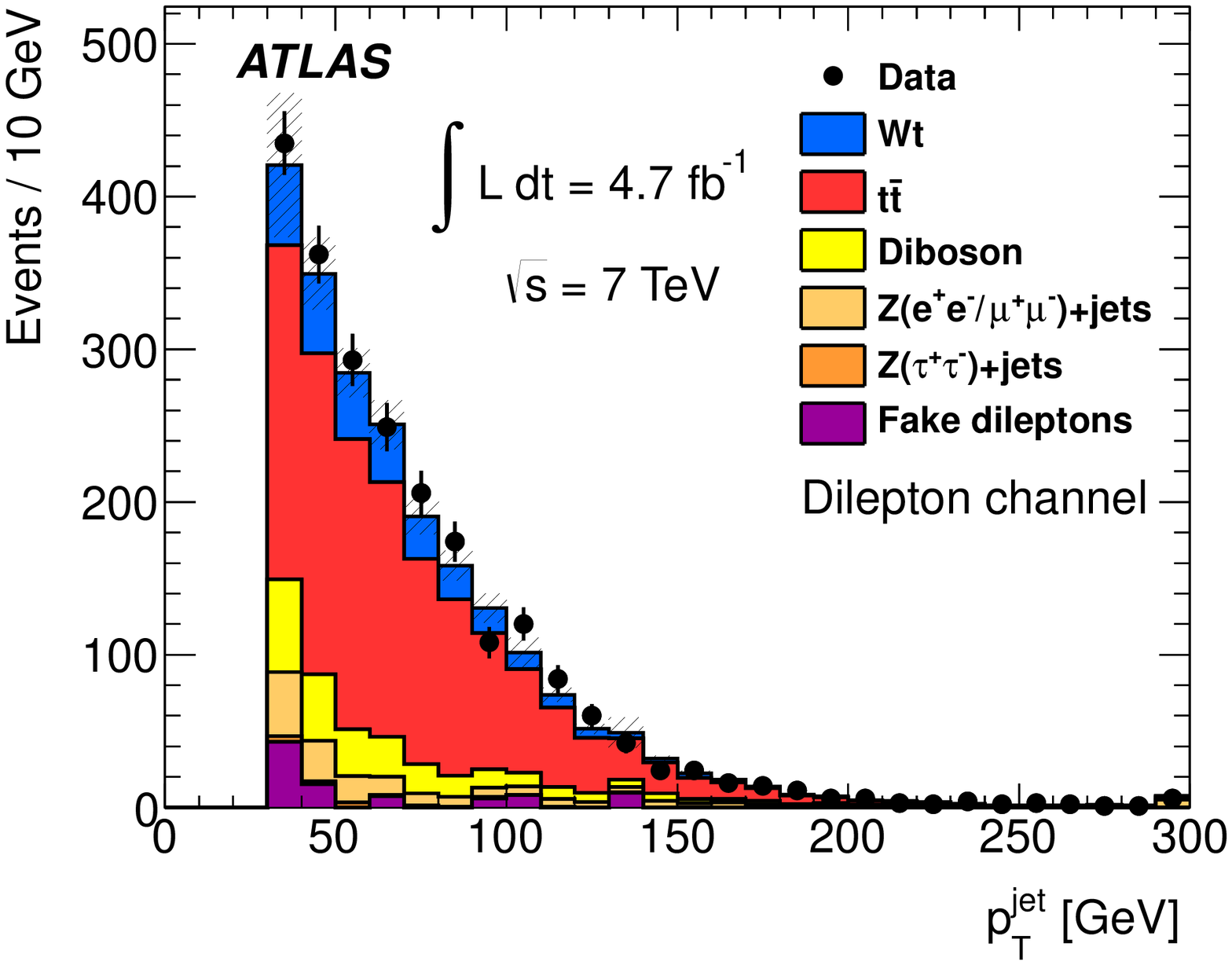}
 \label{fig:ll_jpt}
}
\caption{Kinematic distributions comparing data to predictions in the dilepton channel for
\subref{fig:ll_lpt} the leading lepton $\pt^\mathrm{lep1}$,
\subref{fig:ll_lpt2} the sub-leading lepton $\pt^\mathrm{lep2}$ and
\subref{fig:ll_jpt} the jet $\pt^\mathrm{jet}$.
The hatched band shows the uncertainty due to the background normalisation. 
The last bin includes overflows. }
\label{fig:kinematicsLL1}
\end{figure}

A discriminating variable that separates the signal from the backgrounds is 
$H_\mathrm{T}$, the scalar sum of the transverse momenta of the leptons, jet and
$\met$. The $H_\mathrm{T}$ distribution is shown in Fig.~\ref{fig:ll_ht}.
~
\begin{figure}[!h!tbp]
\centering
\includegraphics[width=0.9\columnwidth]{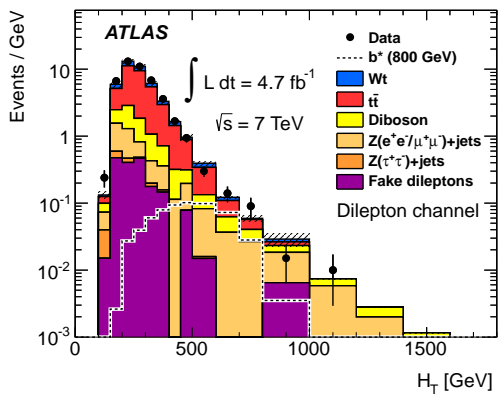}
\caption{$H_\mathrm{T}$ distribution for data and background expectation for the dilepton 
channel.
The hatched band shows the uncertainty due to the background normalisation. 
The signal for a $b^*$-quark mass of $800\GeV$ is also shown.
}
\label{fig:ll_ht}
\end{figure}

\section{Event selection in the lepton+jets channel}
\label{sec:lepjets}
The analysis in the lepton+jets channel follows the same 
background modelling strategy
as the cross-section measurement for single top-quark production in the t-channel~\cite{Aad:2012ux}.
Events are required to have either exactly one muon and $\met >25\GeV$ or exactly one
electron and $\met > 30\GeV$, as well as exactly three jets with $\pt>25\GeV$. Exactly
one of the jets is required to be $b$-tagged to reduce backgrounds. 
The lepton must also match the corresponding trigger object.
Additional requirements are made to reject multijet events, which tend to have low 
\MET\ and a low transverse mass\footnote{The transverse mass, $m_\mathrm{T}^W$, is calculated from the lepton transverse momentum $\pt^\mathrm{lep}$ and the difference of the azimuthal angle, $\Delta \phi$, between the $\met$ and $\pt^\mathrm{lep}$ vector as $m_\mathrm{T}^W = \sqrt{2 \met p_\mathrm{T}^\mathrm{lep} (1-\cos(\Delta \phi(\met,p_\mathrm{T}^\mathrm{lep})))}$} of the lepton--$\met$~system, $m_\mathrm{T}^W$. 
In the muon channel events are required to have $m_\mathrm{T}^W+\MET > 60 \GeV$, while in
the electron channel a requirement of $m_\mathrm{T}^W > 30 \GeV$ is made.
The acceptance for signal events
with $m_{b^*}=800\GeV$ in which one of the $W$~bosons decays leptonically 
($e$ or $\mu$) and the other hadronically is 9\%.

In this channel, one of the largest backgrounds is $W$+jets production for which 
the normalisation and 
flavour composition (the heavy-flavour fraction, HF, includes $b$~quarks and $c$~quarks) 
are derived from data~\cite{Aad:2012qf}.
The overall normalisation is determined from the charge asymmetry between $W^+$ and
$W^-$ production in three-jet events without the $b$-tag requirement.
The flavour composition is determined in two-jet events by comparing the predicted
$W$+jets yields to data with and without a $b$-tag requirement. The resulting 
normalisation and flavour scale factors are then applied to $b$-tagged $W$+3-jets events. 
About 37\% of the total background comes from $W$+jets events, including 28\% from 
events with heavy flavour.

Backgrounds from $\ttbar$ yield 41\% of the total background and single top-quark production in the 
$t$-, $s$- and $Wt$-channel 
9\%.
The multijet background is obtained using a data-based
approach by comparing the numbers of events passing loose and tight
lepton identification criteria~\cite{:2012hg}. It accounts for 9\% of the total 
background. Smaller backgrounds from $Z$+jets and diboson processes are normalised
to their theoretical predictions and contribute 4\%. 

The predicted event yields are compared to data in Table~\ref{tab:lepjetsyield}.
The distributions of the $\pt$ of the highest-$\pt$ jet and $\met$ are shown in
Fig.~\ref{fig:kinematicsLJ}.
~
\begin{table}[!h!tpb]
  \begin{center}
  \caption{Observed and expected event yields in the lepton+jets channel
with normalisation uncertainties. The signal yields are calculated with
$\kappa^b_L=g_L=1$ and $\kappa^b_R=g_R=0$.
\label{tab:lepjetsyield}}
\begin{tabular}{l*{1}{r@{\ $\pm$\ }l}}
 Process
& \multicolumn{2}{c}{Event yield}  \\
\hline \hline
$b^*$ (400 \GeV)  & 12100 &1600 \\
$b^*$ (600 \GeV)  & 1950  &300 \\
$b^*$ (800 \GeV)  & 370   &70  \\
$b^*$ (1000 \GeV) & 79    &17    \\
$b^*$ (1200 \GeV) & 20    &5      \\
\hline
$Wt$                  & 1660                        & 120    \\
single top $s,t$-channel         & 1960                        & 140     \\
$t\bar{t}$            & 15700                       & 1600    \\
$W$+light jets        & 3200                        & 400    \\
$W$+jets HF           & 10900                       & 1400   \\
Diboson               & 327                         & 16     \\
$Z$+jets              & 1300                        & 800     \\
Multijet             & 3500                        & 1700      \\
\hline
Total expected bkg.   & 38500                       & 2900 \\
Total observed        & \multicolumn{2}{c}{38175}\\
\hline\hline
\end{tabular}
\end{center}
\end{table}

\begin{figure}
\subfigure[]{
  \includegraphics[width=0.9\columnwidth]{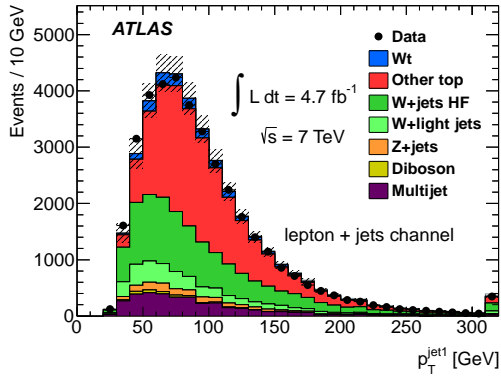}
 \label{fig:lj_jpt}
}
\subfigure[]{
  \includegraphics[width=0.9\columnwidth]{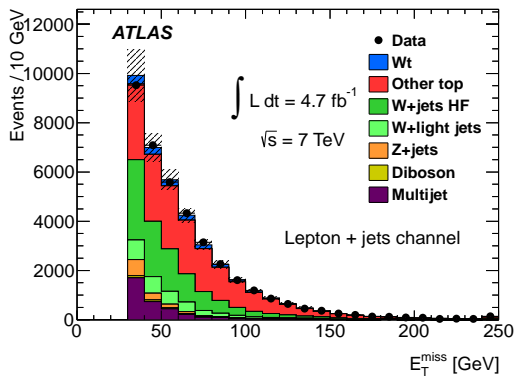}
 \label{fig:lj_met}
}
\caption{Kinematic distributions comparing data to predictions in the lepton+jets 
channel for
\subref{fig:lj_jpt} the $\pt^\mathrm{jet1}$ of the highest-$\pt$ jet and
\subref{fig:lj_met} $\met$. 
``Other top'' includes $\ttbar$, $s$- and $t$-channel single top-quark production.
The hatched band shows the uncertainty due to the background
normalisation. The last bin includes overflows.
}
\label{fig:kinematicsLJ}
\end{figure}

\begin{figure}[!h!tbp]
\includegraphics[width=0.9\columnwidth]{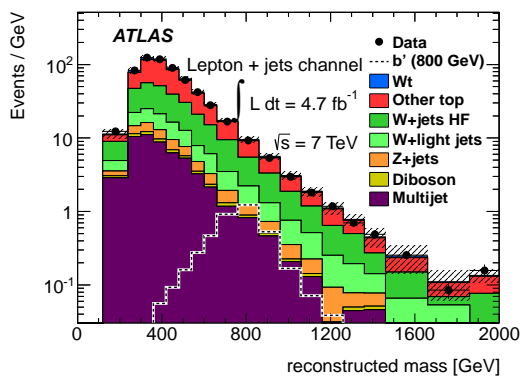}
\caption{
Reconstructed mass distribution for data and background expectation for the 
lepton+jets channel. 
``Other top'' includes $\ttbar$, $s$- and $t$-channel single top-quark production.
The hatched band shows the uncertainty due to the background normalisation. 
The signal for a mass of $800\GeV$ is also shown.
The last bin includes overflows.} 
\label{fig:lj_m}
\end{figure}

In the lepton+jets channel it is possible to reconstruct the candidate 
$b^*$-quark mass from the decay products. The only missing information is the
neutrino longitudinal momentum, which is set to zero. The resulting reconstructed 
mass provides good discrimination between background and signal, as
shown in Fig.~\ref{fig:lj_m}.

\section{Systematic uncertainties}
\label{sec:SYSTEMATICS}
Systematic uncertainties affecting the signal acceptance and the background normalisation
are considered, together with uncertainties affecting the shape of the discriminant
distributions.
The main experimental source of systematic uncertainty comes from the limited 
knowledge of the
jet energy scale~\cite{ATLAS-JES-2011}, which carries an uncertainty of 2--7\% parameterised
as a function of jet $\pt$ and $\eta$. The presence of a $b$~quark in
the jet adds an additional uncertainty of 2--5\% to the jet energy scale uncertainty,
depending on the jet $\pt$.
Other jet-related uncertainty sources are the jet energy resolution, 
jet reconstruction efficiency
and $b$-tagging efficiency~\cite{ATLAS-BTAG}. Lepton-related uncertainties come from trigger and 
identification efficiencies as well as the lepton energy scale and resolution. Event-related 
uncertainties are due to the modelling of multiple proton-proton
interactions and the underlying event as well as $\met$~\cite{ATLAS-MET}. The uncertainty on the
integrated luminosity is $3.9\%$~\cite{ATL-LUMI,ATL-CONF-2011-116}.

Simulation uncertainties include modelling of the hard process, 
parton shower and hadronisation, and initial- and final-state radiation.
These have been
assessed for the $\ttbar$ background events by comparing different generators ({\sc Powheg} and {\sc MC@NLO}),
different shower models ({\sc Pythia} and {\sc Herwig}),
and for $\ttbar$ and signal events different settings for the amount of additional radiation~\cite{ATL-ISR-FSR}.
Other sources of theoretical uncertainty include the normalisation for
$\ttbar$ ($^{+7\%}_{-10\%}$)~\citep{Moch:2008ai,langenfeld:2009tc,Beneke,Aliev:2010zk}, 
single top-quark ($\pm 7\%$)~\cite{Kidonakis:2011wy,Kidonakis:2010ux,Kidonakis:2010tc}
and diboson ($\pm 5\%$ with an additional 24\% per extra jet) production~\cite{Aad:2010ey},
as well as the choice of PDF. The latter was assessed using the
CT10~\cite{Lai:2010vv}, MRST~\cite{springerlink:10} and NNPDF~\cite{Ball:2010de} sets.

Additional uncertainties affect the data-driven background estimation. 
The uncertainty on the DY background normalisation in the dilepton channel is 10\% 
for $ee$ and $\mu\mu$ final 
states and 60\% for $\tau\tau$ final states. The uncertainty on the fake-dileptons 
normalisation in the dilepton channel is 100\%. 
The uncertainty on the $W$+jets normalisation in the lepton+jets channel 
is 13\%. The $W$+jets flavour composition has two additional uncertainties: the
HF contribution has a relative uncertainty of 6\%, 
and the $W_{bb}$/$W_\mathrm{HF}$ ratio has an uncertainty of 17\%.
The multijet background normalisation in
the lepton+jets channel has an uncertainty of 50\%.

\section{Statistical analysis}
\label{sec:limit}
Both the $H_\mathrm{T}$~distribution in the dilepton channel and the reconstructed
mass distribution in the lepton+jets channel show good agreement between the
data and the background model. These two discriminants are used to set limits on the 
$b^*$-quark signal using a Bayesian analysis technique~\cite{Caldwell:2008fw}. The likelihood function is defined as
\begin{equation}
  {\cal L}(\mathrm{data}|\sigma_{b^*}) = \prod_{k} 
  \frac{\mu_k^{n_k} e^{-\mu_k}}{n_k!}\prod_{i}G_i\, ,
\label{eqn:lhoodtemp}
\end{equation}
where $k$ is the index of the discriminant template bin, running over both analysis
channels; $\mu_k = s_k + b_k$ is the sum of predicted signal and background yields;
$n_k$ is the observed yield and
$G_i$ is a Gaussian prior for the $i$th systematic uncertainty.
A flat prior is assumed for the signal cross-section.
Upper limits on the $b^*$-quark production cross-section times branching ratio to $Wt$
are set at the 95\% credibility level (CL) for a series of $b^*$~masses at
$100\GeV$ intervals. 

\begin{figure}[!h!tbp]
\centering
\includegraphics[width=8.0cm]{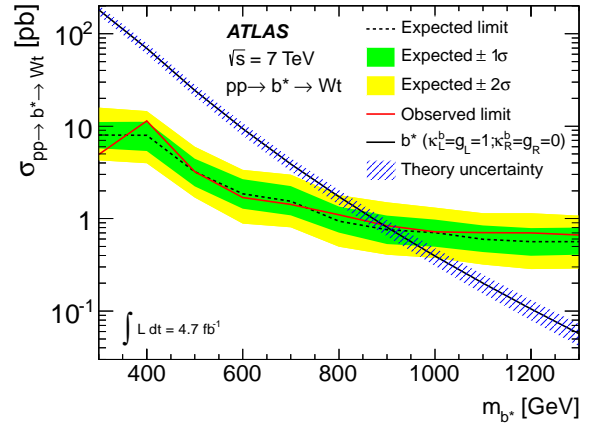}
\caption{Expected and observed limits at the 95\% CL as a function of the $b^*$-quark mass.
Also shown is the theory prediction for $b^*$-quark production with couplings
$\kappa^b_L=g_L=1$ and $\kappa^b_R=g_R=0$, including PDF and scale uncertainties.}
\label{fig:limit}
\end{figure}

The observed and expected cross-section limits as
a function of the $b^*$-quark mass for the left-handed coupling scenario
($\kappa^b_L=g_L=1$ and $\kappa^b_R=g_R=0$) are shown in Fig.~\ref{fig:limit}, where the
expected limit and its uncertainty
are derived from ensembles of background-only pseudo-datasets. The intersection of the 
theoretical cross-section and the observed (expected) cross-section limit defines the
observed (expected) $b^*$-quark mass limit.
The observed lower limit on the $b^*$-quark mass for this left-handed coupling
scenario is $870\GeV$ with an expectation of 
$910\GeV$. When considering only the dilepton channel, the observed (expected) limit on 
the $b^*$-quark mass is $800\GeV$ ($820\GeV$); for the lepton+jets channel, the limits are $800\GeV$ ($830\GeV$).

Limits are also computed for models with right-handed and vector-like couplings of the 
$b^{*}$~quark. 
Setting $\kappa^b_L=g_L=0$ and $\kappa^b_R=g_R=1$,
the observed lower mass limit is $920\GeV$ with an expected limit of $950\GeV$.
Setting $\kappa^b_L=\kappa^b_R=g_L=g_R=1$,
the observed lower mass limit is $1030\GeV$ with an expected limit of $1030\GeV$.

At each mass point, the corresponding cross section is parameterised as a function of the
couplings $\kappa^b_{L,R}$ and
$g_{L,R}$ in order to extract coupling limits in each of the three $b^*$-quark coupling
scenarios.
The resulting limit contours are shown in Fig.~\ref{fig:limit_2d}. The coupling limits
increase as the theoretical cross-section decreases with $b^*$~mass, except for the region
between $400\GeV$ and $500\GeV$ where the backgrounds decrease rapidly with increasing mass
(see Figs.~\ref{fig:ll_ht} and~\ref{fig:lj_m}).

\begin{figure}[!h!tbp]
\centering
\subfigure[]{
\includegraphics[width=8.0cm]{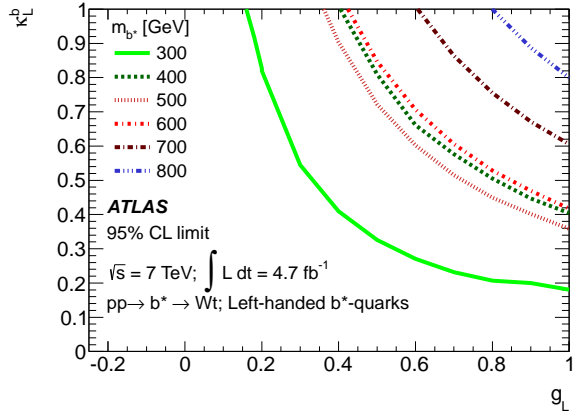}
\label{fig:limit_2dL}
}
\subfigure[]{
\includegraphics[width=8.0cm]{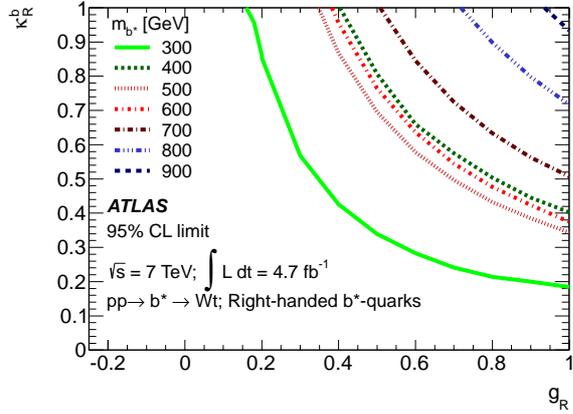}
\label{fig:limit_2dR}
}
\subfigure[]{
\includegraphics[width=8.0cm]{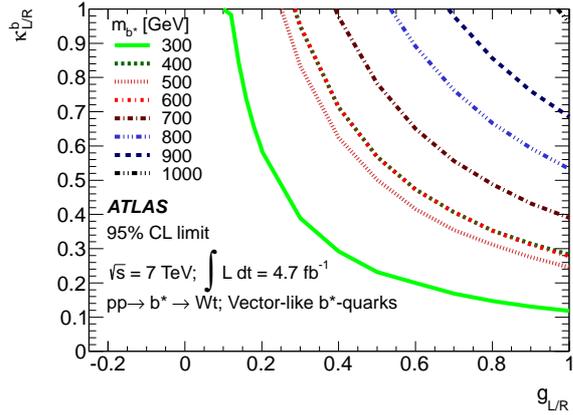}
\label{fig:limit_2dVLQ}
}
\caption{Limit contours at the 95\% CL as a function of the coupling parameters 
for several different $b^*$-quark masses, for
\subref{fig:limit_2dL} left-handed $b^*$~quarks,
\subref{fig:limit_2dR} right-handed $b^*$~quarks and
\subref{fig:limit_2dVLQ} vector-like $b^*$~quarks.
}
\label{fig:limit_2d}
\end{figure}

\section{Summary}
\label{sec:conclusion}
A search for a singly produced excited $b^*$-quark
in $4.7~\ifb$ of data collected with the ATLAS detector in $pp$ collisions at
$\sqrt{s}=7\TeV$ has been presented. 
This is the first search for excited-quarks coupling to the third generation.
It considers the 
dilepton and lepton+jets final states. Limits are computed as a function of the 
$b^* g b$ and $b^* W t$~couplings in three different scenarios.
For purely left-handed couplings and unit strength chromomagnetic coupling, 
$b^*$~quarks with mass below 870 GeV are excluded at the 95\% credibility level.


\input{Acknowledgements}

\bibliographystyle{atlasnote}
\bibliography{dilepton,lj}

\input{atlas_authlist}

\end{document}

%% file: Acknowledgements.tex

\section{Acknowledgements}

We thank CERN for the very successful operation of the LHC, as well as the
support staff from our institutions without whom ATLAS could not be
operated efficiently.

We acknowledge the support of ANPCyT, Argentina; YerPhI, Armenia; ARC,
Australia; BMWF and FWF, Austria; ANAS, Azerbaijan; SSTC, Belarus; CNPq and FAPESP,
Brazil; NSERC, NRC and CFI, Canada; CERN; CONICYT, Chile; CAS, MOST and NSFC,
China; COLCIENCIAS, Colombia; MSMT CR, MPO CR and VSC CR, Czech Republic;
DNRF, DNSRC and Lundbeck Foundation, Denmark; EPLANET, ERC and NSRF, European Union;
IN2P3-CNRS, CEA-DSM/IRFU, France; GNSF, Georgia; BMBF, DFG, HGF, MPG and AvH
Foundation, Germany; GSRT and NSRF, Greece; ISF, MINERVA, GIF, DIP and Benoziyo Center,
Israel; INFN, Italy; MEXT and JSPS, Japan; CNRST, Morocco; FOM and NWO,
Netherlands; BRF and RCN, Norway; MNiSW, Poland; GRICES and FCT, Portugal; MERYS
(MECTS), Romania; MES of Russia and ROSATOM, Russian Federation; JINR; MSTD,
Serbia; MSSR, Slovakia; ARRS and MVZT, Slovenia; DST/NRF, South Africa;
MICINN, Spain; SRC and Wallenberg Foundation, Sweden; SER, SNSF and Cantons of
Bern and Geneva, Switzerland; NSC, Taiwan; TAEK, Turkey; STFC, the Royal
Society and Leverhulme Trust, United Kingdom; DOE and NSF, United States of
America.

The crucial computing support from all WLCG partners is acknowledged
gratefully, in particular from CERN and the ATLAS Tier-1 facilities at
TRIUMF (Canada), NDGF (Denmark, Norway, Sweden), CC-IN2P3 (France),
KIT/GridKA (Germany), INFN-CNAF (Italy), NL-T1 (Netherlands), PIC (Spain),
ASGC (Taiwan), RAL (UK) and BNL (USA) and in the Tier-2 facilities
worldwide.

%% file: atlas_authlist.tex
\onecolumn
\clearpage 
\begin{flushleft}
{\Large The ATLAS Collaboration}

\bigskip

G.~Aad$^{\rm 48}$,
T.~Abajyan$^{\rm 21}$,
B.~Abbott$^{\rm 111}$,
J.~Abdallah$^{\rm 12}$,
S.~Abdel~Khalek$^{\rm 115}$,
A.A.~Abdelalim$^{\rm 49}$,
O.~Abdinov$^{\rm 11}$,
R.~Aben$^{\rm 105}$,
B.~Abi$^{\rm 112}$,
M.~Abolins$^{\rm 88}$
% ...
\bigskip
\\
% ...
$^{11}$ Institute of Physics, Azerbaijan Academy of Sciences, Baku, Azerbaijan\\
$^{12}$ Institut de F{\'\i}sica d'Altes Energies and Departament de F{\'\i}sica de la Universitat Aut{\`o}noma de Barcelona and ICREA, Barcelona, Spain\\
% ...
$^{21}$ Physikalisches Institut, University of Bonn, Bonn, Germany\\
% ...
$^{48}$ Fakult{\"a}t f{\"u}r Mathematik und Physik, Albert-Ludwigs-Universit{\"a}t, Freiburg, Germany\\
$^{49}$ Section de Physique, Universit{\'e} de Gen{\`e}ve, Geneva, Switzerland\\
% ...
$^{88}$ Department of Physics and Astronomy, Michigan State University, East Lansing MI, United States of America\\
% ...
$^{105}$ Nikhef National Institute for Subatomic Physics and University of Amsterdam, Amsterdam, Netherlands\\
% ...
$^{111}$ Homer L. Dodge Department of Physics and Astronomy, University of Oklahoma, Norman OK, United States of America\\
$^{112}$ Department of Physics, Oklahoma State University, Stillwater OK, United States of America\\
% ...
$^{115}$ LAL, Universit{\'e} Paris-Sud and CNRS/IN2P3, Orsay, France\\
% ...
\end{flushleft}